\documentclass[12pt,preprint]{aastex}
\usepackage{graphicx}

\shorttitle{Multiple Main Sequences of very low-mass stars in NGC 2808} 
\shortauthors{A.\ P.\ Milone, et al.\ } 
\usepackage{ulem}

\begin{document}
\title{The infrared eye of the Wide-Field Camera 3 on the {\it Hubble Space Telescope} reveals multiple main sequences of very low-mass stars in NGC 2808.
          \footnote{           Based on observations with  the
                               NASA/ESA {\it Hubble Space Telescope},
                               obtained at  the Space Telescope Science
                               Institute,  which is operated by AURA, Inc.,
                               under NASA contract NAS 5-26555.}}

\author{
A.\ P. \,Milone\altaffilmark{2,3},
A.\ F. \,Marino\altaffilmark{4}, 
S.\ Cassisi\altaffilmark{5},     
G.\ Piotto\altaffilmark{6,7},
L.\ R. \,Bedin\altaffilmark{7},
J.\ Anderson\altaffilmark{8}, 
F.\ Allard\altaffilmark{9}, 
A.\ Aparicio\altaffilmark{2,3}, 
A.\ Bellini\altaffilmark{8},
R.\ Buonanno\altaffilmark{5,10},
M.\ Monelli\altaffilmark{2,3},
A.\ Pietrinferni\altaffilmark{5}
 }

\altaffiltext{2}{Instituto de Astrof\`\i sica de Canarias, E-38200 La
              Laguna, Tenerife, Canary Islands, Spain; [milone, aparicio, monelli]@iac.es}

\altaffiltext{3}{Department of Astrophysics, University of La Laguna,
           E-38200 La Laguna, Tenerife, Canary Islands, Spain}

\altaffiltext{4}{Max-Planck-Institut f\"{u}r Astrophysik, Karl-Schwarzschild-Str. 1, 85741 Garching bei M\"{u}nchen, Germany; amarino@MPA-Garching.MPG.DE}

\altaffiltext{5}{INAF-Osservatorio Astronomico di Collurania, via Mentore
           Maggini, I-64100 Teramo, Italy [cassisi, pietrinferni]@oa-teramo.inaf.it} 

\altaffiltext{6}{Dipartimento  di   Astronomia,  Universit\`a  di Padova,
           Vicolo dell'Osservatorio 3, Padova I-35122, Italy;
           giampaolo.piotto@unipd.it }

\altaffiltext{7}{INAF-Osservatorio Astronomico di Padova, Vicolo
             dell'Osservatorio 5, I-35122 Padova, Italy;
             luigi.bedin@oapd.inaf.it}

\altaffiltext{8}{Space Telescope Science Institute,
                  3800 San Martin Drive, Baltimore,
                  MD 21218; [jayander,bellini]@stsci.edu}

\altaffiltext{9} {CRAL, UMR 5574, CNRS, Universit\`e de Lyon, \`Ecole Normale Sup\`erieure de Lyon, 46 All\`ee d'Italie, F-69364 Lyon Cedex 07, France; france.allard@me.com}

\altaffiltext{10} {Dipartimento di Fisica, Universit\'a di Roma Tor Vergata, Via della Ricerca Scientifica 1, 00133 Rome, Italy;  buonanno@roma2.infn.it }

\begin{abstract}
We use images taken with the infrared channel of the Wide Field Camera 3 on 
the {\it Hubble Space Telescope (HST)} to study the multiple main sequences (MSs) 
of NGC~2808. Below the turn off, the red, the middle, and the blue MS, 
previously detected from visual-band
photometry, are visible over an interval of about 3.5 F160W magnitudes. 
The three MSs merge together at the level of the MS bend. At fainter 
magnitudes, the MS again splits
into two components containing $\sim$65\% and $\sim$35\% of stars, with 
the most-populated MS being the bluest one.  
Theoretical isochrones suggest that the latter is connected to the 
red MS discovered in the optical color-magnitude diagram (CMD), and 
hence corresponds to the first stellar generation, having primordial 
helium and enhanced carbon and oxygen abundances.  
The less-populated MS in the faint part of the near-IR CMD  is 
helium-rich and poor in carbon and oxygen, 
and it can be associated with the middle and the blue MS of the 
optical CMD.  The finding that the photometric
signature of abundance anticorrelation are also present in fully convective 
MS stars reinforces the inference that they have a primordial origin.
\end{abstract}


\section{Introduction}
\label{introduction}

In recent years, photometric studies have shown that the
color-magnitude diagrams (CMDs) of globular clusters (GCs) can be
very complex, with the presence of multiple main sequences (MSs,
e.\ g.\ Anderson 1997, Bedin et al.\ 2004, Piotto et al.\ 2007, hereafter P07),
multiple sub-giant branches (SGBs, e.\ g.\ Milone et al.\ 2008 and 2012a,
Anderson et al.\ 2009, Piotto et al.\ 2012), and multiple or spread
red-giant branches (RGBs, e.\ g.\ Yong et al.\ 2008, Marino et
al.\ 2008, Lee et al.\ 2011).

Photometric and spectroscopic investigations have revealed that 
the multiple sequences in the CMDs of many GCs are 
populated by stars with different helium and light-element
abundances (e.\ g.\ Piotto et al.\ 2005, 2007, Yong et al.\ 2008, 
Marino et al.\ 2008, Sbordone et al.\ 2011, hereafter S11).  Is some cases, the presence 
of stellar populations with different composition (in particular, 
Helium) has been associated with the presence of multimodal or extended 
horizontal-branches (HBs) (e.\ g.\ D'Antona et al.\ 2005, P07, 
D'Antona \& Caloi 2008,  Marino et al.\ 2011).  

Among clusters with multiple stellar populations, NGC~2808 is certainly
one of the most intriguing objects.
Its CMD shows a multimodal MS (D'Antona et al.\ 2005) composed of three distinct components (P07, Milone et al.\ 2012b, hereafter M12), 
a multimodal HB, which is greatly extended blueward (Sosin et al.\ 1997, 
Bedin et al.\ 2000), and a spread RGB (Lee et al.\ 2011).
Furthermore, spectroscopic studies of RGB, HB, and bright MS stars have 
revealed significant star-to-star variations in the light-element 
abundances, with three distinct groups of stars populating an extended 
Na-O anti-correlation (Carretta et al.\ 2006, Gratton et al.\ 2011, 
Bragaglia et al.\ 2010).  

Photometric studies, based on data collected with the Wide Field Channel 
of the Advanced Camera for Survey (WFC/ACS) on board {\it HST} have made
it possible to detect and characterize the multiple MSs of NGC~2808 
from the MSTO down to about 4 magnitudes below the MSTO (P07, M12): 
the stars all have almost the same age and [Fe/H] but different Helium
and light-element abundances.
The red MS (rMS) corresponds to a first generation
and has primordial helium and light-element abundances (Y$\sim$0.25), 
while the blue (bMS) and the middle (mMS) 
include later generations of stars, and are both enhanced in 
helium, sodium, and nitrogen (Y$\sim$0.38 and Y$\sim$0.32, respectively) 
and depleted in oxygen and carbon (D'Antona et al.\ 2005, P07).

Usually, photometry of GC sequences extends over a 
limited spectral region, from the ultra-violet ($\lambda \sim$ 2000\AA) 
to the near-infrared (NIR, $\lambda \sim$ 8000\AA).  As such, multiple
sequences are rarely detected along the lower part of the MS, because 
observational limits make it hard to get high-accuracy photometry of 
very faint and red stars in optical and UV colors. 

In this Letter we use {\it HST} to extend the study to the
near-infrared passbands  and complement the work by S11 who analysed the signatures of the anticorrelations in the visual and ultra-violect portion of the spectrum.   
We analyze the CMD of NGC~2808 through the F110W ($\sim${\it J}) and F160W
($\sim${\it H}) filters of the infra-red channel of the Wide Field
Camera 3 (WFC3/IR) and follow, for the first time, the multiple
sequences of this cluster over a wide interval of stellar masses, 
from the turn off down to very low-mass (VLM) MS stars 
($\mathcal{M} \sim$0.2$\mathcal{M}_{\odot}$).  

\section{Observations and data reduction}
\label{data}
In this work we used archival {\it HST} images taken with the WFC3/IR 
camera for program GO-11665 (P.\ I.\ Brown).  This data set consists 
of 2$\times$699s exposures through F110W and two through F160W of 799s 
and 899s for each of three different WFC3/NIR fields, taken in parallel
while STIS was taking spectra of EHB stars.  All fields are located at 
about six arcmin from the cluster center.  
 
In the following, fields `A', `B', and `C' will correspond, respectively, 
to the fields in the: south-west, south, and north-east of the cluster 
center.  Field `A' partially overlaps the ACS/WFC field analyzed by 
P07 and M12. Therefore, for a sub-sample of the measured stars, we have
both visual and NIR photometry and can perform a direct comparison of 
the multi-band photometry (see Sect.~\ref{results}).

The images were reduced by using a software package that is based largely
on the algorithms described by Anderson \& King (2006) and will be presented in a separate paper. Star positions are corrected for geometric distortion
by using the solution given by Anderson et al.\ (in preparation), and
were calibrated as in Bedin et al.\ (2005) using the current on-line 
estimates for zero points and encircled energies\footnote{ %
       \textsf{http:\/\/www.stsci.edu\/hst\/wfc3\/phot\_zp\_lbn}
}.

The analysis we present here requires high-precision photometry, so we 
selected a high-quality sample of stars that (1) have a good fit to 
the PSF, (2) are relatively isolated, (3) and have small astrometric 
and photometric errors (see Milone et al.\ 2009, Sect.~2.1 for details
of this procedure).  Finally we corrected our photometry for differential 
reddening using the method that is described in great detail in Milone 
et al.\ (2012c).  Briefly, the method consists in defining a MS
ridge line in the CMD.  Then, for each star, we selected the 35 nearest 
well-measured neighbors and evaluated for each of them the color distance 
from the fiducial line along the reddening line.  We applied to the target
star a correction equal to the median color distance of these 35 stars.   

\section{The NIR Color-Magnitude diagram of NGC\,2808}
\label{results}
\begin{figure}[ht!]
   \centering
   \epsscale{.75}
   \plotone{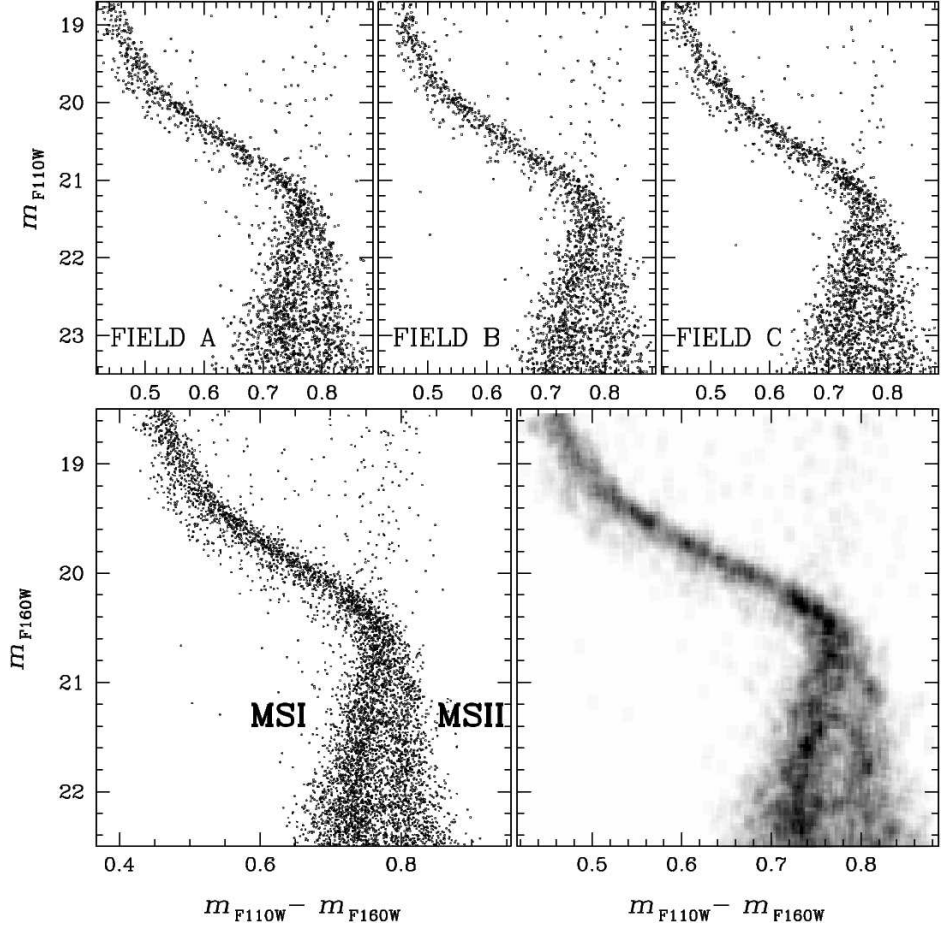}
   \caption{\textit{Upper panels:} 
                    $m_{\rm F110W}$ versus $m_{\rm F110W}-m_{\rm F160W}$ 
                    CMD for stars in the fields `A', `B', and `C' 
                    corrected for differential reddening. 
            \textit{Lower panels:} 
                    $m_{\rm F160W}$ versus $m_{\rm F110W}-m_{\rm F160W}$ 
                    CMD (left) and Hess diagram (right) for all the stars. 
            }
   \label{fig1}
   \end{figure}

The IR $m_{\rm F110W}$ versus $m_{\rm F110W}-m_{\rm F160W}$ CMDs of stars in the fields `A', `B' and `C' are plotted in the
upper panels of Fig.~\ref{fig1}, while the lower-left and lower-right
panels show the $m_{\rm F160W}$ versus $m_{\rm F110W}-m_{\rm F160W}$ CMD and the Hess diagram for all the stars. 
The most obvious feature in these diagrams is the multi-modal
MS that stretches from the turn-off (just below saturation) to 
 about three F160W magnitudes below it.
Furthermore, the MS's breadth and color distribution dramatically 
changes when moving from bright to faint magnitudes.  Down to 
$m_{\rm F160W} \sim 20.5$, the color distribution across the MSs is 
consistent with what has been previously observed in the optical.  
As expected from theory, below this, the MS runs almost vertically
(actually slightly shifting towards bluer colors), and the three 
MS components discovered by P07 appear to merge at this bending point. 
However, below $m_{\rm  F160W}\sim 21$, the MS split re-appears, 
but this time with a different morphology.  Whereas in the upper part 
of the CMD, and as in the optical bands, the less-populated MS is the 
bluest component, below  $m_{\rm  F160W}\sim 21$, the less-populated 
component is the reddest one.  In the following, we will refer to these 
two MS-components fainter than $m_{\rm  F160W}\sim 21$, as $MS_{\rm I}$ 
(more-populated and bluer), and $MS_{\rm II}$ (less-populated MS and 
redder). 

\begin{figure}[ht!]
   \centering
   \epsscale{.75}
   \plotone{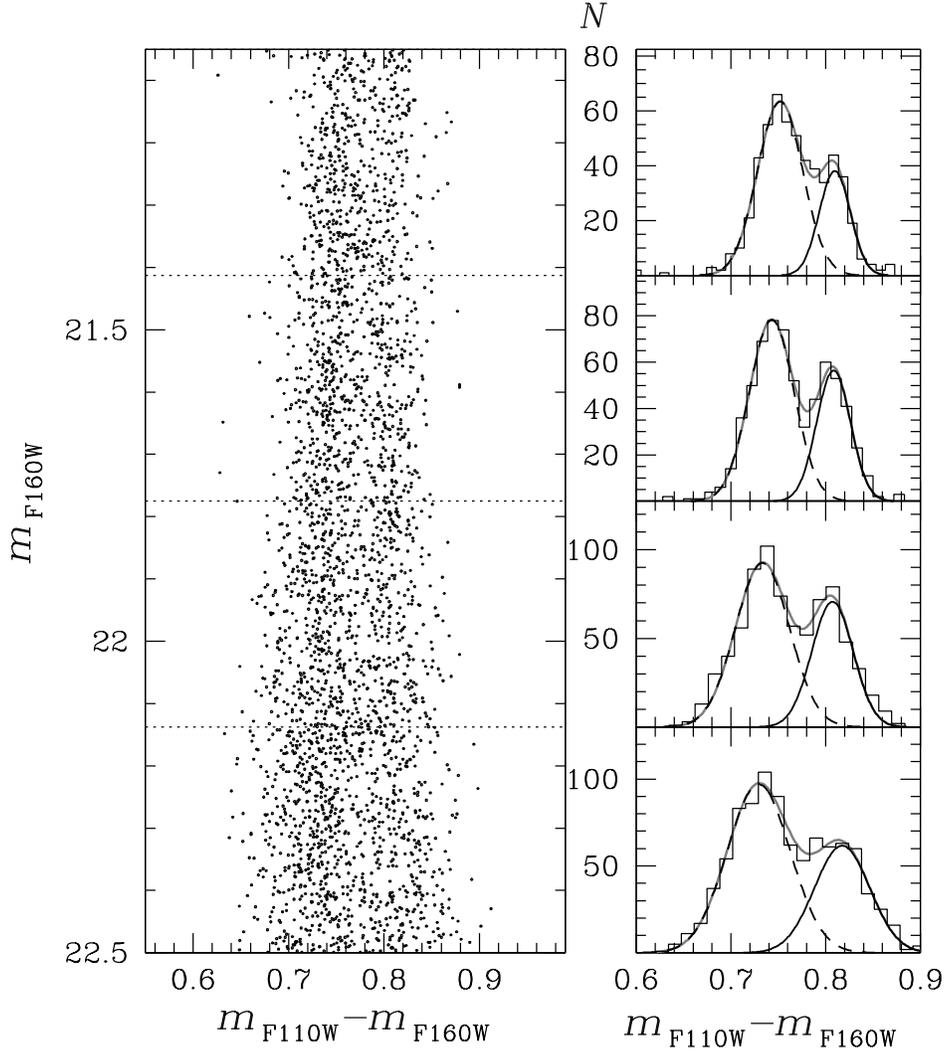}
   \caption{\textit{Left panel:}  
                    A close-up of the region of the $m_{\rm F160W}$ 
                    versus $m_{\rm F110W}-m_{\rm F160W}$ CMD from 
                    the lower-left panel of Fig.~1, where the MS split is more evident. 
            \textit{Right panel:} 
                    The color distribution of the stars plotted in the
                    left panel, in four F160W magnitude intervals. 
                    The continuous gray lines are fits by a sum of 
                    two Gaussians.
            }
   \label{fig2}
   \end{figure}

To estimate the fraction of stars in each MS we followed the procedure 
illustrated in Fig.~\ref{fig2}, which is similar to that used in several 
previous papers (e.g.\ P07).  The left panel shows a zoom-in of the 
$m_{\rm F160W}$ versus $m_{\rm F110W}-m_{\rm F160W}$ CMD region where 
the MS split is most evident.  We analyze the color distribution of the 
stars plotted in the left panel in four magnitude intervals over the range 
$21.25<m_{\rm F160W}<22.5$.  The distribution is clearly bimodal, and has 
been fitted with two Gaussians (dashed and continuous black lines for the $MS_{\rm I}$ and 
$MS_{\rm II}$, respectively).  From the areas under the Gaussians, 
we find that 65$\pm$2\% of stars belong to the $MS_{\rm I}$, and 35$\pm$2\%
to the $MS_{\rm II}$.  The errors are calculated as the rms of the 
measurements obtained in the $N$=4 magnitude intervals divided by $\sqrt{N-1}$.
 
\begin{figure}[ht!]
   \centering
   \epsscale{.75}
   \plotone{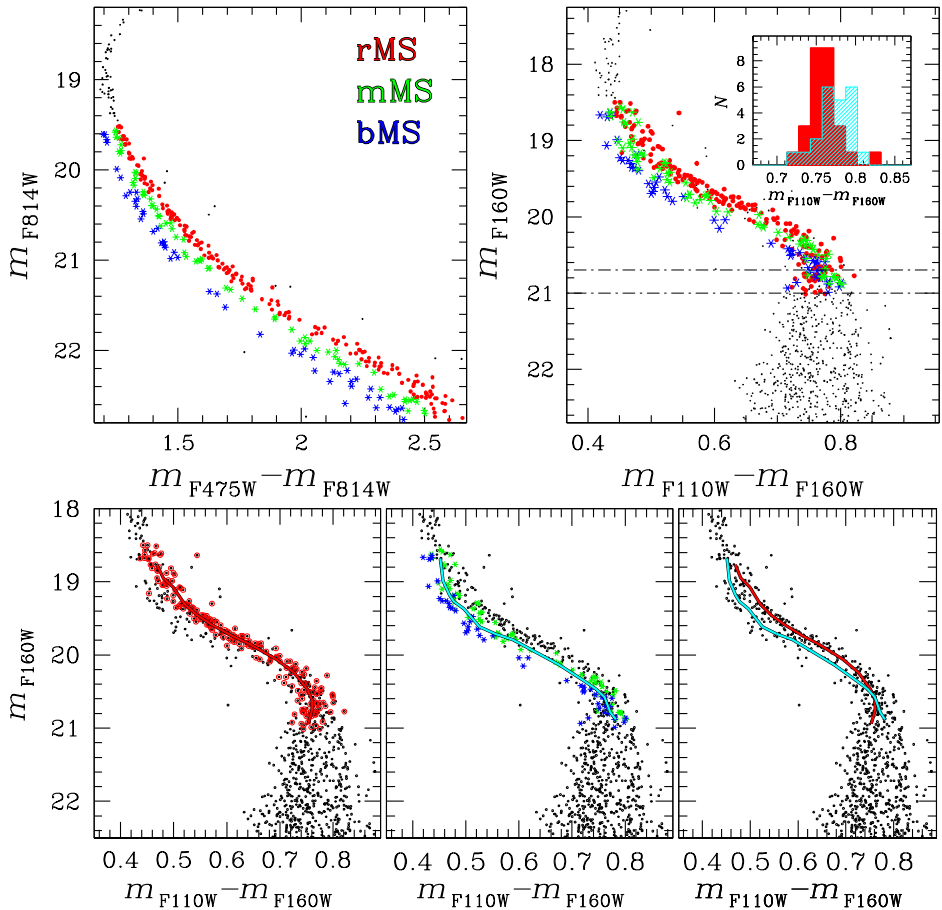}
   \caption{\textit{Upper panels:} 
                    $m_{\rm F814W}$  versus $m_{\rm F475W}-m_{\rm F814W}$ 
                    (left) and $m_{\rm F160W}$ versus 
                    $m_{\rm F110W}-m_{\rm F160W}$ CMD for stars in 
                    field `A' (right). The rMS, mMS, and bMS stars are 
                    colored red, green, and blue, respectively. 
                    The color distribution for MS stars between the two 
                    dashed-dotted lines is shown in the inset where red 
                    and cyan histograms correspond to rMS stars and to 
                    the sample of both mMS and bMS stars respectively 
                    (see text for details). 
            \textit{Lower-panel:} 
                    Fiducial line for the rMS stars (left) and for stars 
                    in the other two MSs (middle) superimposed on the 
                    IR CMD.  A comparison of the two fiducial lines is 
                    shown in the right panel.  
           }
   \label{fig3}
   \end{figure}

We note that the fractions of stars along $MS_{\rm I}$ and $MS_{\rm II}$ 
are very similar to the fraction of rMS (62$\pm$2\%) stars and the total 
fraction of mMS and bMS stars (24+14=38$\pm$3\%, M12).  This fact makes 
it very tempting to associate the $MS_{\rm I}$ with the rMS (defined 
by P07), while both their mMS and their bMS could be the extension 
to bright magnitudes of the $MS_{\rm II}$.  To further investigate this 
issue, in Fig.~\ref{fig3}, we show photometry for the stars in field `A'
that also happen to be measured in P07.  The three groups bMS, mMS, and 
rMS defined in P07 are colored in blue, green, and red, and are plotted 
with the same color-code in the other panels.  In the upper-right panel, 
we show the $m_{\rm F160W}$ vs.\ $m_{\rm F110W}-m_{\rm F160W}$ CMD for 
the same stars cross-identified in the WFC3/NIR bandpasses.  We note 
that:  
  i) \textit{above} the MS bend ($m_{\rm F160W} \simeq 20.5$) the three MSs 
      exhibit in the NIR the same relative locations as in the optical CMD,  
 ii) \textit{At} the magnitude of the MS bend, the three MSs appear to merge,
      and
iii) \textit{Below} the MS bend, the bMS and mMS appear to become redder 
      than the stars labeled as rMS in the optical photometry. 
The histograms of the NIR-color-distribution in the inset of the top-right 
panel of Fig.~3 appear to confirm this trend.  The cyan and the red histograms show 
the color distribution for  bMS+mMS stars and rMS respectively in the magnitude interval 20.75$<m_{\rm F160W}<$21.00
The cyan histogram (bMS+mMS) has, on average,
bluer colors than red histogram (rMS). 

In the lower-left panel we show the NIR fiducial lines for the optically 
selected rMS sample of stars (in red).  Defining this fiducial line is 
an iterative procedure.  We started from our optically defined sample 
of rMS stars (from P07 and M12), then we calculated the sigma-clipped 
median of NIR colors and of NIR magnitudes for rMS stars in intervals 
of 0.2 mag in F160W, and interpolated with a spline.  Similarly, we 
obtained the fiducial lines for the combined sample of mMS and bMS stars 
shown in the middle panel.  The two fiducial lines are compared in the 
lower-right panel and confirm that the rMS is indeed redder than the 
mMS and bMS for magnitudes fainter than  $m_{\rm  F160W} \sim$ 20.6.  
It is therefore legitimate to associate the rMS with the component 
MS$_{\rm I}$, and bMS+mMS with the MS$_{\rm II}$.  

\section{Discussion}
The near-IR/WFC3 CMD presented in this Letter allows us to examine,
for the first time, the behavior of multiple MSs among VLM stars. 
As expected, in the near-IR, the MS stars less massive than 
$\sim$0.4$\mathcal\, M_{\odot}$ define a sequence with nearly-constant 
color  (see e.g.\ Baraffe et al.\ 1997, Zoccali et al.\ 2000, Bono et al.\ 2010 and 
references therein).  Theoretical models predict that this is due 
to two competing effects.  On one hand, the increase of the radiative 
opacity, coupled with  the decrease of the effective temperature shifts 
the stellar colors to the red.  On the other hand, the increase of the 
collisional induced absorption (CIA) of the H$_{2}$ molecule in the 
infrared moves back the stellar flux to the blue.  In a given range 
of stellar masses, the two effects compensate, and the MS locus runs 
almost vertical.  Moving towards less massive stars the second effect 
becomes dominant, and the color of the MS becomes bluer and bluer with 
decreasing stellar mass.  The CIA source is obviously related to the abundance 
of H$_2$ molecules. It also scales as the square of the density ($\rho$), 
as opposed to other opacity sources, which are proportional to $\rho$ 
(Cassisi 2011 and references therein).  

In order to compare empirical evidence with suitable evolutionary predictions, we have computed some grids
of evolutionary models for both low- and very low-mass stars. For the low-mass structures (i.\ e.\ ${\rm \mathcal{M}>0.5\mathcal{M}_\odot}$) we
adopt the physical scenario described in Pietrinferni et al.\ (2006), while in the VLM regime we use the same physical inputs adopted in Cassisi et
al.\ (2000). We address the interested reader to the quoted references for details.  The match between the more massive models and the VLM
ones was made at a mass level where the transition in luminosity and effective temperature between the two regimes is smooth (usually ${\rm \sim 0.5\mathcal{M}_\odot}$). We computed stellar models for an iron content equal to ${\rm [Fe/H]=-1.3}$, and $\alpha-$elements enhancement
equal to ${\rm [\alpha/Fe]=+0.4}$, and for three values for the initial He abundances: Y=0.248, 0.35 and 0.40, namely. These initial He abundances have been selected on the basis of the comparison between stellar models and the triple MS observed in the optical CMD of NGC~2808 performed by P07.

The theoretical models have been transformed  into the observational domain by integrating the synthetic spectra of the 
BT-Settl AGSS model atmosphere grid\footnote{These model atmospheres and colors tables are available via the 
Phoenix web simulator at the following URL site: http://phoenix.ens-lyon.fr/simulator.} (Allard et al.\ 2011, 2012) over the IR WFC3 bandpasses. We used the filter transmission tables provided by the System Throughputs  web site for the WFC3.  

\begin{figure}[ht!]
   \centering
   \epsscale{.99}
   \plotone{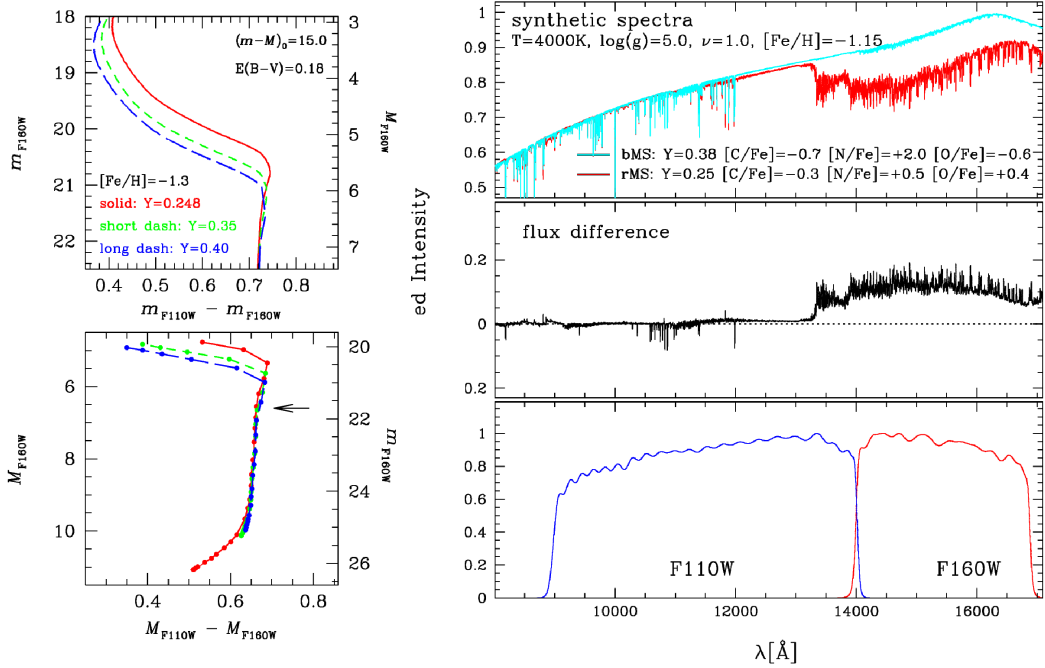}
   \caption{\textit{Left panels}: theoretical isochrones for an age of
                    12~Gyr, a metallicity suitable for NGC~2808 and 
                    various assumptions about the initial He content 
                    (see labels). 
                    The upper-left panel shows the theoretical absolute 
                        magnitudes and colors. 
                    In the lower-left panel, we show -after applying a distance 
                        modulus $\rm (m-M)_{0}$=15.0 and a reddening 
                        E(B$-$V)=0.18, in order to make the comparison 
                        with the upper right panel of Fig.~3 easier- 
                      the location in the CMD of the
                    stellar models with mass 
                    in the range from ${\rm 0.0894M_{\odot}}$ to 
                    ${\rm 0.65M_{\odot}}$.  
                     The arrow marks the approximate 
                    location along the MS loci of the VLM stellar model 
                    with effective temperature and gravity values 
                    consistent with those adopted for computing the model 
                    atmospheres.
            \textit{Right panels}:  Comparison of the synthetic spectra
                    of an bMS star (cyan) and an rMS star (red, see text 
                    for more details) difference between the synthetic 
                    spectrum of a  bMS and a  rMS star (middle). 
                    Normalized responses of the F110W and F160W WFC3/NIR {\it HST} 
                    filters (bottom).
  }
         \label{fig4}
   \end{figure}

The upper-left panel of Fig.~\ref{fig4} shows the location in the selected near-IR  CMD of 12-Gyr isochrones for the selected assumptions about the initial He abundances.  Data in Fig.~\ref{fig4} show that for $m_{\rm F160W} \lesssim 21$ (i.\ e.\ $M_{\rm F160W}\lesssim6$, the He abundance plays a fundamental role in driving  the MS location: when increasing the He abundance from Y=0.248 to 0.40 the MS locus becomes fainter by more than 0.5~mag at fixed color. This is consistent (and indeed explains) the separation of the three MSs observed in the upper part ($m_{\rm F160W}<20.5$) of the 
CMDs of Figs.~1 and 3.  However,  since helium-rich stars have lower hydrogen abundance, we would expect that, in the stellar mass regime where the two - previously quoted - physical processes are in competition,  the He-rich VLM stars would appear redder than He-normal stars for their lower H$_{2}$ abundance, and consequent lower contribution coming from the H$_2$ CIA.  Indeed, at fainter magnitudes, the He-normal (Y=0.248) MS sequence runs on  the blue side of the more He-rich sequences, but the effect is very small, smaller than the observed one.

 We note that, below the faint limit of our photometry ($M_{\rm F160W}\lesssim10$), the He-normal sequence reaches fainter magnitudes and bluer colors than the He-rich counterpart.
This is an important feature that could be tested with deeper CMDs, and may represent an independent  observational confirmation of high-He enhancement.

The bottom-left panel of Fig.~\ref{fig4} shows that the difference in the initial He abundances among the stars belonging to the distinct MS loci
is not able to provide a complete explanation of the observed trend.  In the following, we will attempt to identify the origin of the (small) 
separation of the observed MSs at magnitudes fainter than 
$M_{\rm F160W}\sim 20.5$.  First of all, we note that the 
model atmospheres adopted for computing the color-${\rm T_{eff}}$ transformations do not account for the peculiar chemical patterns  of the various sub-populations present in NGC~2808.  In particular, they do not take into account the effect of light-elements anti-correlations, which have been shown to be relevant for the  MS color in the UV and Str{\"o}mgren bands (S11, M12). 

In order to explore the impact of light-element variations on the 
NIR WFC3 bands, we have computed synthetic spectra trying to account 
for the chemical patterns of the stars belonging to the distinct 
MSs in NGC~2808.  For this calculation, we used the ATLAS9 and SYNTHE Kurucz programs in the range from {8000\AA}\, to {18000\AA}\, (Kurucz 2005, Sbordone et al. 2007)\footnote{\sf{http://wwwuser.oat.ts.astro.it/castelli/}}.  

For the $MS_{\rm I}$ and $MS_{\rm II}$, we adopted a Helium content of Y=0.25 and Y=0.38 respectively, as suggested by isochrone-fitting on the upper MS (M12).  
For the $MS_{\rm I}$, we assumed average chemical abundance 
of O-rich stars ([O/Fe]=0.4, as measured by Carretta et al.\ 2006), 
and adopted the carbon and nitrogen abundance ([C/Fe]=$-$0.3, [N/Fe]=0.5) 
measured by Bragaglia et al.\ (2010).  For the $MS_{\rm II}$, we used 
 [O/Fe]=$-$0.6 (the O abundance measured by Carretta 
et al.\ 2006 for the O-poor group), and [C/Fe]=$-$0.7, and [N/Fe]=2.0 
(as measured by Bragaglia et al.\ 2010 for a bMS star).
Table~1 summarizes the adopted chemical abundances for the two sequences.    
For both MSs we used  ${\rm T_{\rm eff}=4000K}$, log(g)=5.0, and a microturbolence 1.0 $\rm {km~s^{-1}}$. 
Our synthesis includes the following molecules in the  Kurucz compilation: 
$\rm {CO}$, $\rm {C_{2}}$, $\rm {CN}$, $\rm {OH}$, $\rm {MgH}$, $\rm {SiH}$, $\rm {H_{2}O}$, $\rm {TiO}$ ($\rm {H_{2}O}$, from Partridge \& Schwenke 1997; $\rm {TiO}$ from Schwenke 1998) $\rm {VO}$ and $\rm {ZrO}$ (B.\ Plez priv. communication)
The resulting synthetic spectra (Fig.~4, upper-right panel) have been integrated over the transmission of the WFC3/IR F110W and the
F160W filters (lower-right panel) to produce synthetic magnitudes and colors. 
We found a $\Delta(M_{\rm F110W}-M_{\rm F160W})=0.10$ color difference between 
the simulated $MS_{\rm I}$ and $MS_{\rm II}$ stars, consistent 
with the observed color difference (Fig.~2) at $m_{\rm F160W}$=21.5: 
$\Delta(m_{\rm F110W}-m_{\rm F160W})$=0.06$\pm$0.01 mag. 
The middle-right panel of Fig.~4 shows the flux difference as a function
of the wavelength between the two MSs spectra.
The $\rm {H_{2}O}$  molecules have the strongest effect on the synthetic spectra, and cause the significant lower flux of the rMS in the filter F160W.
We consider the agreement between the simulated 
and observed color differences satisfactory, accounting for the high 
sensitivity to light-elements abundances of the stellar spectra at these 
wavelengths, and that the adopted C and N abundances are based on the 
measurement on just one bMS and one rMS star (Bragaglia et al.\ 2010).

\begin{table}
\center
\begin{tabular}{ccccc}
\hline
\hline
MS   &  Y    & [C/Fe] & [N/Fe]   & [O/Fe] \\
\hline
rMS & 0.25  & $-$0.3   &  0.5     &   0.4 \\
mMS & 0.32  &  NA    &  NA      &   0.0 \\
bMS & 0.38  & $-$0.7   &  2.0     &  $-$0.6 \\
\hline
MSI & 0.25  & $-$0.3   &  0.5     &   0.4 \\ 
MSII& 0.38  & $-$0.7   &  2.0     &  $-$0.6 \\ 
\hline
\hline
\end{tabular}
\label{tab1}
\caption{Average chemical abundances of bMS, mMS, and rMS stars
and abundanced adopted for the $MS_{\rm I}$ and the $MS_{\rm II}$.}
\end{table} 
\section{Summary}
The photometry presented in this paper for NGC~2808 reveals, for the first time, multiple sequences in near-infrared CMD of a GC. The brightest part of the CMD 
 ($m_{\rm F160W}<20.5$) is consistent with three populations with different He and light-element abundance, as already noticed in previous papers based 
on visual photometry. The three MSs merge together at the luminosity of the MS
bend while 
at fainter magnitudes, a combination of stellar structure 
and atmospheric effects makes the distinction among the different stellar 
populations more intricate.
Our CMD allows us to identify at least two 
MSs.  A redder, more populated $MS_{\rm I}$, which includes $\sim65$\% 
of the MS stars, and that we associate with the first stellar generation,
which has primordial He, and O-C-rich/N-poor stars, and a  $MS_{\rm II}$, 
with $\sim35$\% of stars, corresponding to a second generation stellar 
population that is enriched in He and N and depleted in C and O.  
The $MS_{\rm I}$ of Fig.~1 is the faint counterpart of the rMS identified 
by P07, whereas the $MS_{\rm II}$ corresponds to the lower mass 
counterpart of the mMS and bMS of P07. 

 This Letter provides the first detection of multiple populations with different helium and light-element abundances among VLM stars and extends the investigation by S11 to the near-IR. The fact that the signatures of abundance anticorrelation are also observed among fully-convective M-dwarfs demonstrates, once and for all, that they have primordial origin and hence correspond to different stellar generations. 
 
\begin{acknowledgements}
We are gratefull to the referee for a thoughtful, thorough report that improved the accuracy and focus of the paper.
We warmly thank B.\ Plez for kindly providing molecular linelists.
Support for this work has been provided by the IAC (grant 310394), and the Education and Science Ministry of Spain (grants AYA2007-3E3506, and AYA2010-16717).
S.C.\ and G.P.\ thanks for financial support from PRIN INAF "Formation and early evolution of massive star clusters". 
G.P.\ acknowledges support by ASI under grants ASI-INAF I/016/07/0 and I/009/10/0.
\end{acknowledgements}
%
%
\bibliographystyle{aa}

\begin{thebibliography}{}

\bibitem[Allard et al.(2011)]{2011arXiv1112.3591A} Allard, F., Homeier, D., 
\& Freytag, B.\ 2011, arXiv:1112.3591 

\bibitem[Allard et al.(2012)]{} Allard, F., Homeier, D., Freytag, B.\ 2012, Philosophical Transactions A, in press.

\bibitem[Anderson(1997)]{1997PhDT.........8A} Anderson, A.~J.\ 1997, Ph.D.~Thesis 


\bibitem[Anderson \& King(2006)]{2006acs..rept....1A} Anderson, J., \& King, I.~R.\ 2006, Instrument Science Report ACS 2006-01, 34 pages, 1 

\bibitem[Anderson et al.(2009)]{2009ApJ...697L..58A} Anderson, J., Piotto, G., King, I.~R., Bedin, L.~R., \& Guhathakurta, P.\ 2009, \apjl, 697, L58 

\bibitem[Baraffe et al.(1997)]{1997A&A...327.1054B} Baraffe, I., Chabrier, G., Allard, F., \& Hauschildt, P.~H.\ 1997, \aap, 327, 1054 

\bibitem[Bedin et al.(2000)]{2000A&A...363..159B} Bedin, L.~R., Piotto, G., Zoccali, M., et al.\ 2000, \aap, 363, 159 

\bibitem[Bedin et al.(2004)]{2004ApJ...605L.125B} Bedin, L.~R., Piotto, G., Anderson, J., et al.\ 2004, \apjl, 605, L125 

\bibitem[Bedin et al.(2005)]{2005MNRAS.357.1038B} Bedin, L.~R., Cassisi, S., Castelli, F., et al.\ 2005, \mnras, 357, 1038 

\bibitem[Bono et al.(2010)]{2010ApJ...708L..74B} Bono, G., Stetson, P.~B., VandenBerg, D.~A., et al.\ 2010, \apjl, 708, L74 

\bibitem[Bragaglia et al.(2010)]{2010ApJ...720L..41B} Bragaglia, A., Carretta, E., Gratton, R.~G., et al.\ 2010, \apjl, 720, L41 

\bibitem[Calamida et al.(2009)]{2009IAUS..258..189C} Calamida, A., Bono, G., Stetson, P.~B., et al.\ 2009, IAU Symposium, 258, 189 

\bibitem[Carretta et al.(2006)]{2006A&A...450..523C} Carretta, E., Bragaglia, A., Gratton, R.~G., et al.\ 2006, \aap, 450, 523 

\bibitem[Cassisi et al.(2000)]{2000MNRAS.315..679C} Cassisi, S., 
Castellani, V., Ciarcelluti, P., Piotto, G., \& Zoccali, M.\ 2000, \mnras, 315, 679 

\bibitem[Cassisi(2011)]{2011arXiv1111.6464C} Cassisi, S.\ 2011, arXiv:1111.6464 

\bibitem[D'Antona et al.(2005)]{2005ApJ...631..868D} D'Antona, F., Bellazzini, M., Caloi, V., et al.\ 2005, \apj, 631, 868 

\bibitem[D'Antona \& Caloi(2008)]{2008MNRAS.390..693D} D'Antona, F., \& Caloi, V.\ 2008, \mnras, 390, 693 

\bibitem[Gratton et al.(2011)]{2011A&A...534A.123G} Gratton, R.~G., Lucatello, S., Carretta, E., et al.\ 2011, \aap, 534, A123 

\bibitem[Kurucz(2005)]{2005MSAIS...8...14K} Kurucz, R.~L.\ 2005, Memorie della Societa Astronomica Italiana Supplementi, 8, 14 

\bibitem[Lee et al.(2009)]{2009Natur.462..480L} Lee, J.-W., Kang, Y.-W., Lee, J., \& Lee, Y.-W.\ 2009, \nat, 462, 480 

\bibitem[Marino et al.(2008)]{2008A&A...490..625M} Marino, A.~F., Villanova, S., Piotto, G., et al.\ 2008, \aap, 490, 625 

\bibitem[Marino et al.(2011)]{2011ApJ...730L..16M} Marino, A.~F., Villanova, S., Milone, A.~P., et al.\ 2011, \apjl, 730, L16 

\bibitem[Milone et al.(2008)]{2008ApJ...673..241M} Milone, A.~P., Bedin, L.~R., Piotto, G., et al.\ 2008, \apj, 673, 241 

\bibitem[Milone et al.(2009)]{2009A&A...497..755M} Milone, A.~P., Bedin, L.~R., Piotto, G., \& Anderson, J.\ 2009, \aap, 497, 755 

\bibitem[Milone et al.(2012)]{2012ApJ...744...58M} Milone, A.~P., Piotto, 
G., Bedin, L.~R., et al.\ 2012a, \apj, 744, 58 

\bibitem[Milone et al.(2012)]{2012ApJ...744...58M} Milone, A.~P., Piotto, G., Bedin, L.~R., et al.\ 2012b, \apj, 744, 58, M12 

\bibitem[Milone et al.(2012)]{2012A&A...540A..16M} Milone, A.~P., Piotto, G., Bedin, L.~R., et al.\ 2012c, \aap, 540, A16 

\bibitem[Norris(2004)]{2004ApJ...612L..25N} Norris, J.~E.\ 2004, \apjl, 612, L25 
\bibitem[Partridge(1997)]{PS1997} Partridge \& Schwenke(1997)Partridge and Schwenke Partridge, H., \& Schwenke, D. W. 1997, J. Chem. Phys., 106, 4618.

\bibitem[Pietrinferni et al.(2006)]{2006ApJ...642..797P} Pietrinferni, A., 
Cassisi, S., Salaris, M., \& Castelli, F.\ 2006, \apj, 642, 797 

\bibitem[Piotto et al.(2005)]{2005ApJ...621..777P} Piotto, G., et al.\ 2005, \apj, 621, 777 

\bibitem[Piotto et al.(2007)]{2007ApJ...661L..53P} Piotto, G., Bedin, L.~R., Anderson, J., et al.\ 2007, \apjl, 661, L53 , P07

\bibitem[Piotto et al.(2012)]{} Piotto, G., Milone, A.~P., Bedin, L.~R., et al.\ 2012, submitted to \apj

\bibitem[Sbordone et al.(2007)]{2007IAUS..239...71S} Sbordone, L., Bonifacio, P., \& Castelli, F.\ 2007, IAU Symposium, 239, 71 

\bibitem[Sbordone et al.(2011)]{2011A&A...534A...9S} Sbordone, L., Salaris, M., Weiss, A., \& Cassisi, S.\ 2011, \aap, 534, A9, S11 

\bibitem[Schwenke(1998)]{1998FaDi..109..321S} Schwenke, D.~W.\ 1998, 
Faraday Discussions, 109, 321 

\bibitem[Sosin et al.(1997)]{1997ApJ...480L..35S} Sosin, C., Dorman, B., 
Djorgovski, S.~G., et al.\ 1997, \apjl, 480, L35 

\bibitem[Yong et al.(2008)]{2008ApJ...684.1159Y} Yong, D., Grundahl, F., Johnson, J.~A., \& Asplund, M.\ 2008, \apj, 684, 1159 

\bibitem[Zoccali et al.(2000)]{zoccali00} Zoccali, M., Cassisi, S., Frogel, J.A., Gould, A., Ortolani, S., Renzini, A., Rich, R. M., Stephens, A.W. 2000, \apj, 530, 418

\end{thebibliography}

\end{document}